\documentclass[fleqn,twoside]{article}
\usepackage{espcrc2}

\usepackage{graphicx}
\usepackage{epsfig}
\usepackage[figuresright]{rotating}
\usepackage{bm}
\usepackage{amssymb}

\newcommand{\AmS}{{\protect\the\textfont2
  A\kern-.1667em\lower.5ex\hbox{M}\kern-.125emS}}

\title{Status of the Hadronic $\tau$ Decay Determination of 
$\vert V_{us}\vert$}

\author{K. Maltman$^\dagger$\thanks{$^\dagger$Alternate address: CSSM, 
University of Adelaide, Adelaide
SA 5005 Australia}, C.E. Wolfe
\address{Dept. Mathematics and Statistics, York Univ.,
4700 Keele St., Toronto, ON CANADA M3J 1P3}; %
     S. Banerjee, I.M. Nugent, and J.M. Roney
\address{Dept. Physics and Astronomy, Univ. Victoria,
PO Box 3055, Victoria, BC CANADA V8W 3P6}}
       
\begin{document}

\begin{abstract}
We update the hadronic $\tau$ determination of 
$\vert V_{us}\vert$, showing that current strange branching
fractions produce results $2-3\sigma$ lower than 3-family unitarity
expectations. Issues related to the size of theoretical
uncertainties and results from an alternate, mixed $\tau$-electroproduction 
sum rule determination are also considered.
\vspace{1pc}
\end{abstract}

\maketitle

\section{Introduction and Background}
The determination of $\vert V_{us}\vert$ from hadronic $\tau$ decay data 
rests on the finite energy sum rule (FESR) relation, 
\begin{equation}
\int_0^{s_0}w(s)\, \rho(s)\, ds\, =\, -{\frac{1}{2\pi i}}\oint_{\vert
s\vert =s_0}w(s)\, \Pi (s)\, ds
\label{basicfesr}
\end{equation}
valid for any analytic $w(s)$ and kinematic-singularity-free correlator, 
$\Pi$, having spectral function, $\rho (s)$. To obtain $\vert V_{us}\vert$,
Eq.~(\ref{basicfesr}) is applied to the flavor-breaking (FB) correlator 
difference $\Delta\Pi_\tau (s)\, \equiv\,
\left[ \Pi_{V+A;ud}^{(0+1)}(s)\, -\, \Pi_{V+A;us}^{(0+1)}(s)\right]$,
where $\Pi^{(J)}_{V/A;ij}$ are the spin $J=0,1$ components
of the flavor $ij$, vector (V) or axial vector (A) current two-point
functions, and $(0+1)$ denotes the sum of $J=0$ and $1$ components.
The OPE is to be employed on the RHS for sufficiently large $s_0$.

The spectral functions, $\rho^{(J)}_{V/A;ij}$, 
are related to the differential distributions, $dR_{V/A;ij}/ds$,
of the normalized flavor $ij$ V or A current induced decay widths,
$R_{V/A;ij}\, \equiv\, \Gamma [\tau^- \rightarrow \nu_\tau
\, {\rm hadrons}_{V/A;ij}\, (\gamma )]/ \Gamma [\tau^- \rightarrow
\nu_\tau e^- {\bar \nu}_e (\gamma)]$, by~\cite{tsai}
\begin{eqnarray}
&&{\frac{dR_{V/A;ij}}{ds}}\, =\, c^{EW}_\tau \vert V_{ij}\vert^2
\left[ w_{L+T}^{(00)}(y_\tau ) \rho_{V/A;ij}^{(0+1)}(s)
\right.\nonumber\\
&&\left.\ \ \  
- w_L^{(00)}(y_\tau )\rho_{V/A;ij}^{(0)}(s) \right]
\label{basictaudecay}\end{eqnarray}
with $y_\tau =s/m_\tau^2$, $w_{L+T}^{(00)}(y)=(1-y)^2(1+2y)$,
$w_L^{(00)}(y)=2y(1-y)^2$, $V_{ij}$ the flavor $ij$ CKM matrix element,
and, with $S_{EW}$ a short-distance electroweak correction~\cite{erler02}, 
$c^{EW}_\tau \equiv 12\pi^2 S_{EW}/m_\tau^2$.

Use of the $J=0+1$, FB difference $\Delta\Pi_\tau$,
rather than the analogous difference involving the linear combination of
$J=0,0+1$ spectral functions appearing in Eq.~(\ref{basictaudecay}),
is a consequence of the extremely bad behavior of the integrated 
$J=0$ (longitudinal) $D=2$ OPE series~\cite{longprob}. 
Fortunately, apart from the accurately known $\pi$ and $K$ pole terms,
contributions to $\rho_{V+A;ij}^{(0)}$ are $\propto [(m_i\mp m_j)^2]$, 
making $ud$ continuum contributions negligible. Once the small continuum 
$us$ $J=0$ contributions are determined phenomenologically using 
dispersive~\cite{jop} and sum rule~\cite{mksps} analyses of the 
strange scalar and pseudoscalar channels, respectively, the $J=0$
contributions can be subtracted, bin-by-bin, from $dR_{V+A;ij}/ds$,
allowing one to construct the re-weighted $J=0+1$ spectral integrals,
$R^w_{V+A;ij}(s_0)$, defined by
\begin{equation}
{\frac{R^w_{V+A;ij}(s_0)}{c^{EW}_\tau \vert V_{ij}\vert^2}}\,
\equiv\, \int_0^{s_0}ds\, w(s)\, \rho^{(0+1)}_{V+A;ij}(s)\ ,
\end{equation}
and, from these, the FB combinations,
\begin{eqnarray}
\delta R^w_{V+A}(s_0)\, =\,&&
{\frac{R^w_{V+A;ud}(s_0)}{\vert V_{ud}\vert^2}}
\, -\, {\frac{R^w_{V+A;us}(s_0)}{\vert V_{us}\vert^2}}\nonumber\\
=\,&&c^{EW}_\tau\, \int_0^{s_0}ds\, w(s)\Delta\rho_\tau (s)\ .
\label{tauvusbasicidea}\end{eqnarray}
With $\vert V_{ud}\vert$, and any parameters in
$\delta R_{V+A}^{w,OPE}(s_0)
=c^{EW}_\tau\left[{\frac{-1}{2\pi i}}\oint_{\vert s\vert = s_0}
ds\, w(s)\Delta\Pi_\tau (s)\right]$ from 
other sources, Eq.~(\ref{basicfesr}) yields~\cite{gamizetal}
\begin{equation}
\vert V_{us}\vert \, =\, \sqrt{{\frac{R^w_{V+A;us}(s_0)}
{{\frac{R^w_{V+A;ud}(s_0)}{\vert V_{ud}\vert^2}}
\, -\, \delta R^{w,OPE}_{V+A}(s_0)}}}\ .
\label{tauvussolution}\end{equation}
$\delta R^{w,OPE}_{V+A}(s_0)$ is typically $<< R^w_{V+A;ud,us}(s_0)$ 
(usually at the few-to-several-$\%$ 
level) for $s_0\gtrsim 2\ {\rm GeV}^2$, 
making a high precision $\vert V_{us}\vert$ determination
possible with only modest OPE precision~\cite{gamizetal}.

It turns out (see also below) that the convergence of the
integrated $J=0+1$, $D=2$ OPE series may also be somewhat problematic.
As a result, it is also of interest to consider FESRs based on
the alternate FB correlator difference, 
\begin{equation}
\Delta\Pi_M\equiv 9\Pi_{EM}-5\Pi^{(0+1)}_{V;ud}
+\Pi^{(0+1)}_{A;ud}-\Pi^{(0+1)}_{V+A;us} ,\label{altcorr}
\end{equation}
where $\Pi_{EM}$ is the scalar part of the electromagnetic (EM) current
two-point function. $\Delta\Pi_M$ shares with $\Delta\Pi_\tau$ the vanishing 
of $D=0$ contributions to all orders but, by construction, has strongly
suppressed $D=2$ contributions~\cite{kmtauem08}.
$D=4$ contributions turn out also strongly suppressed compared
to those of $\Delta\Pi_\tau$. This suppression 
does not, however, persist beyond $D=4$~\cite{kmtauem08}. 
The EM spectral function, $\rho_{EM}(s)$, required on the LHS of
the $\Delta\Pi_M$ FESR, is given by
$\rho_{EM}(s)={\frac{s\sigma_0(s)}{16\pi^2\alpha_{EM}^2}}$, with
$\sigma_0(s)$ the bare inclusive hadronic electroproduction cross-section.
The $\Delta\Pi_M$ FESR yields a solution for $\vert V_{us}\vert$
of the form Eq.~(\ref{tauvussolution}), with
the RHS denominator replaced by 
$9R^w_{EM}(s_0) -5{\frac{R^w_{V;ud}(s_0)}{\vert V_{ud}\vert^2}}
+{\frac{R^w_{A;ud}(s_0)}{\vert V_{ud}\vert^2}}-\delta R_{w,M}^{OPE}(s_0)$,
where $R^w_{EM}(s_0)=c^{EW}_\tau\, \int_0^{s_0}ds\, w(s)\rho_{EM}(s)$
and
\begin{equation}
\delta R_{w,M}^{OPE}(s_0)={\frac{-c^{EW}_\tau}{2\pi i}}
\, \oint_{\vert s\vert = s_0}ds\, w(s)\Delta\Pi_M (s)\, .
\end{equation}

\section{\label{sec2}Spectral and OPE Input}
\subsection{Spectral Input}
We compute $R_{V+A;ud}^w(s_0)$ and $R^w_{V+A;us}(s_0)$ using
the publicly available ALEPH $us$~\cite{alephus99} and $ud$~\cite{alephud05} 
spectral data and covariances. Separate $ud$ V and A
analogues, $R_{V/A;ud}^w(s_0)$, required for the mixed 
$\tau$-electroproduction FESRs implement the improved 
$\bar{K}K\pi$ V/A $ud$ separation~\cite{davieretal08}
provided by CVC and the BaBar determination of $I=1$ $K\bar{K}\pi$
electroproduction cross-sections~\cite{babarkkbarpi07}. 
A small global rescaling of the continuum $ud$ V+A distribution 
accounts for recent changes in $S_{EW}$, $R_{V+A;us}$ and $B_e$. 
We employ $\vert V_{ud}\vert =0.97425(23)$~\cite{hardytowner} and
current values~\cite{banerjee08} for $B_e$, $R_{V+A;us}$ and $R_{V+A;ud}$.
Since BaBar and Belle have not yet completed their remeasurements of 
$dR_{V+A;us}/ds$, we work with an interim partial 
update obtained by rescaling the 1999 ALEPH distribution~\cite{alephus99} 
mode-by-mode by the ratio of new to old world averages for the
branching fractions~\cite{alephrescale}. The new world averages, based on
the results of
Refs.~\cite{babar07a,babar07b,babar08a,babar08b,belle06,belle07a,belle08a,belle08b},
are given in Table~\ref{table1}~\cite{banerjee08}.
The $us$ V+A covariance matrix cannot yet be analogously updated,
so the improved precision on the $us$ branching fractions
translates into an improved $us$ spectral integral error 
only for $w=w_{L+T}^{(00)}\equiv w_{(00)}$ and $s_0=m_\tau^2$.

Details of the treatment of the EM spectral data, required for the 
spectral integral side of the $\Delta\Pi_M$ FESR, are
omitted here because of space constraints, but may be found in 
Ref.~\cite{kmtauem08}.

\begin{table}[th]
\caption{Current world averages, $B[\tau^-\rightarrow X^-_{us}\nu_\tau]$ 
for the main strange hadronic modes $X^-_{us}$, 
with $S$ factors where required. Column 3 gives the references
associated with post-PDG2006 changes.}
{\begin{tabular}{lll}
\hline
$X^-_{us}$&${\cal{B}}_{WA,2008}\, (\%)$&Refs.\\
\hline
$K^-$ [$\tau$ decay]&$0.690(10)$&\cite{banerjee08,babar08b} \\
\ \ \ \ \ \ ([$K_{\mu 2}$])&($0.715(4)$)& \\
$K^-\pi^0$&$0.426(16)$&\cite{belle07a}\\
$\bar{K}^0\pi^-$&$0.835(22)$ ($S=1.4$)&\cite{babar08a,belle08b}\\
$K^-\pi^0\pi^0$&$0.058(24)$&\\
$\bar{K}^0\pi^0\pi^-$&$0.360(40)$& \\
$K^-\pi^-\pi^+$&$0.290(18)$ ($S=2.3$)&\cite{babar07b,belle08a}\\
$K^-\eta$&$0.016(1)$&\cite{belle08b}\\
$(\bar{K}3\pi )^-$ (est'd)&$0.074(30)$& \\
$K^-\omega$&$0.067(21)$&\\
$(\bar{K}4\pi )^-$ (est'd)&$0.011(7)$&\\
$K^{*-}\eta$&$0.014(1)$&\cite{belle08b} \\
$K^-\phi$&$0.0037(3)$ ($S=1.3$)&\cite{babar07b,belle06} \\
\hline
TOTAL&$2.845(69)$&\\
&($2.870(68)$)&\\
\hline
\end{tabular} \label{table1}}
\end{table}

\begin{figure}
  \caption{\label{fig1} $\vert V_{us}\vert$ versus $s_0$ from
the $\Delta\Pi_\tau$ FESRs for, from top to bottom,
$w_{20}$, $\hat{w}_{10}$, $w_{10}$ and $w_{(00)}$.}
\rotatebox{270}{\mbox{
  \begin{minipage}[t]{2.6in}
\includegraphics[width=2.55in]{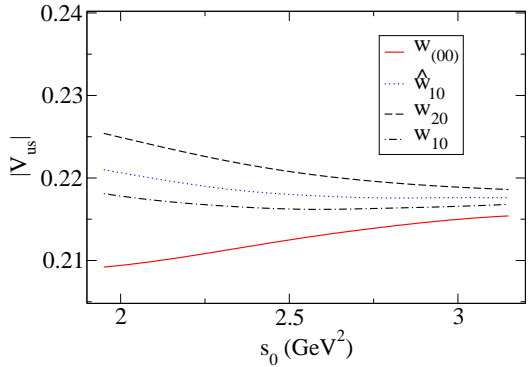}
  \end{minipage}
}}
\end{figure}

\subsection{OPE input}
To keep OPE-breaking contributions from the vicinity of the timelike
point on $\vert s\vert =s_0$ sufficiently suppressed, we restrict
our attention to $w(s)$ having at least a double zero at $s=s_0$,
and to $s_0\gtrsim 2\ {\rm GeV^2}$~\cite{opecontrol}.

The leading, $D=2$, OPE contribution to $\Delta\Pi_\tau$
is known to 4 loops~\cite{d2ope}:
\begin{eqnarray}
&&\left[\Delta\Pi_\tau (Q^2)\right]^{OPE}_{D=2}\, =\, {\frac{3}{2\pi^2}}\,
{\frac{m_s(Q^2)}{Q^2}} \left[ 1\, +\, {\frac{7}{3}} \bar{a}\, \right.\nonumber
\\
&&\left. \qquad +\, 19.93 \bar{a}^2 \, +\, 208.75 \bar{a}^3
\, +\, \cdots \right]\ ,
\label{d2form}\end{eqnarray}
with $\bar{a}=\alpha_s(Q^2)/\pi$, and $\alpha_s(Q^2)$ and $m_s(Q^2)$
the running coupling and strange quark mass in the $\overline{MS}$ scheme.
Since independent determinations of $\alpha_s$ imply 
$\bar{a}(m_\tau^2)\simeq 0.1$, convergence at the spacelike point on 
$\vert s\vert = s_0$ is marginal at best. With such slow 
convergence, conventional prescriptions for assessing the $D=2$ truncation 
uncertainty may lead to significant underestimates. 

To deal with the potential $D=2$ convergence problem, one 
may either work with $\Delta\Pi_\tau$ and
$w(s)$ chosen to emphasize regions of the complex $s\, =\, -Q^2$-plane away 
from the spacelike point, where $\vert \alpha_s (Q^2)\vert$ is smaller
and convergence improved~\cite{km00}, or switch to the alternate 
$\Delta\Pi_M$ FESRs where $D=2$ contributions are
suppressed already at the correlator level~\cite{kmtauem08}. 
In the latter case, the $D=2$ contribution becomes
\begin{eqnarray}
&&\left[\Delta\Pi_{\tau , EM} (Q^2)\right]^{OPE}_{D=2}\, 
=\, {\frac{3}{2\pi^2}}\,
{\frac{m_s(Q^2)}{Q^2}} \left[ \  {\frac{1}{3}} \bar{a}\, \right.\nonumber\\
&&\left. \qquad +\, 4.384 \bar{a}^2 \, +\, 44.94 \bar{a}^3
\, +\, \cdots \right]\ ,
\label{d2altform}\end{eqnarray}
more than an order of magnitude smaller than in the $\Delta\Pi_\tau$ case.
Since $\alpha_s(s_0)$ grows with decreasing $s_0$, making higher order terms 
relatively more important at lower scales, extracted $\vert V_{us}\vert$ 
results will display an unphysical $s_0$-dependence if neglected, higher order 
$D=2$ terms are, in fact, not negligible. $s_0$-stability studies thus
provide a handle on the impact of the potentially slow integrated
$D=2$ convergence on $\vert V_{us}\vert$.

$D=4$ OPE contributions to $\Delta\Pi_\tau (Q^2)$
and $\Delta\Pi_M (Q^2)$ are determined by
$\langle m_s\bar{s}s\rangle$ and $\langle m_\ell\bar{\ell}\ell\rangle$, up to
negligible $O(m_s^4)$ corrections. The relevant expressions, as 
well as those for the $D=6$ four-quark condensate contributions, are easily 
constructed from the results of Ref.~\cite{bnp}, and given
in Ref.~\cite{kmtauem08}. If one works
with weights $w(s)=\sum_{m=0}b_my^m$, with $y=s/s_0$, integrated $D=2k+2$
OPE terms scale as $1/s_0^k$, allowing contributions of different
$D$ to be distinguished by their differing $s_0$-dependences. 

As $D=2$ OPE input, we employ 
$m_s(2\ {\rm GeV})=96(10)$ MeV~\cite{gamizetalnew} and
$\alpha_s(m_\tau^2)=0.323(9)$, the latter obtained from
an average, $\alpha_s(M_Z)=0.1190(10)$, of various recent determinations
(including lattice~\cite{newlattice} and $\tau$~\cite{alphatau08} results,
which are now in very good agreement)
via the standard combination of 4-loop running and 3-loop matching at the 
flavor thresholds~\cite{cks97}.

At $D=4$, we employ the GMOR relation for 
$\langle m_\ell\bar{\ell}\ell\rangle$ and evaluate $\langle m_s\bar{s}s\rangle$
using the ChPT determination of $m_s/m_\ell$~\cite{leutwylermq}
and $\langle m_s\bar{s}s\rangle /\langle m_\ell\bar{\ell}\ell\rangle =1.2(3)$,
the latter obtained by updating Ref.~\cite{jaminssoverll} using
the average of recent $n_f=2+1$ lattice determinations of $f_{B_s}/f_B$
as input~\cite{latticefbsoverfb}. 

$D=6$ contributions are estimated
using the vacuum saturation approximation (VSA), rescaled by 
$\rho_{VSA}=1(5)$, while $D>6$ contributions are neglected.
Since integrated $D\geq 6$ OPE contributions scale as $1/s_0^N$ ($N\geq 2$), 
if $D>4$ contributions are, in fact, not small, 
and these input assumptions are unreliable, 
an unphysical $s_0$-dependence of $\vert V_{us}\vert$
will result, again making $s_0$-stability tests important.

\begin{figure}
  \caption{\label{fig2} $\vert V_{us}\vert$ versus $s_0$ from
the $\Delta\Pi_\tau$ FESRs for, from top to bottom,
$w_{20}$, $\hat{w}_{10}$, $w_{10}$ and $w_{(00)}$, with the spectral
input modified by rescaling up by $3\sigma$ the branching fraction of the
large, but not yet remeasured, $\bar{K}^0\pi^-\pi^0$ mode.}
\rotatebox{270}{\mbox{
  \begin{minipage}[t]{2.6in}
\includegraphics[width=2.55in]{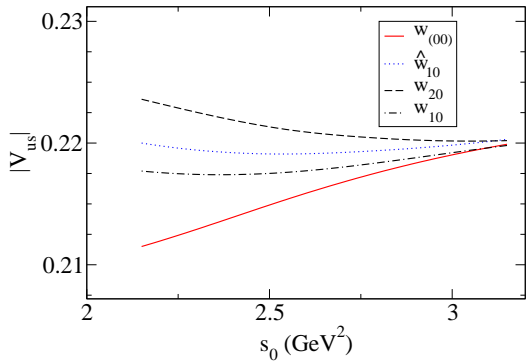}
  \end{minipage}
}}
\end{figure}

\section{Results and discussion}
The results for $\vert V_{us}\vert$ obtained using the inputs 
specified above for the $\Delta\Pi_\tau$ FESRs based on the 
$J=0+1$ kinematic weight $w_{(00)}(y)$,
and three weights, $w_{10}(y)$, $w_{20}(y)$, $\hat{w}_{10}(y)$,
constructed in Ref.~\cite{km00} specifically to improve the
poor integrated $J=0+1$, $D=2$ convergence, are displayed in 
Fig.~\ref{fig1}. The $s_0$-instability of the $w_{(00)}$
results is much greater than the theoretical uncertainty
$\sim \pm 0.0005$ often quoted for the $s_0=m_\tau^2$
version of this analysis in the literature.
The results corresponding to $\hat{w}_{10}$, in contrast, 
display a very good window of $s_0$-stability. A positive
feature of the $\Delta\Pi_\tau$ analysis is the fact that the results 
for all four weights appear to be converging towards the stable $\hat{w}_{10}$ 
value as $s_0\rightarrow m_\tau^2$. The $s_0=m_\tau^2$ versions of the
various analyses are 
\begin{equation}
\vert V_{us}\vert = \left\{
\begin{array}{ll}
0.2180(32)(15)& (\hat{w}_{10})\\
0.2188(29)(22)& (w_{20})\\
0.2172(34)(11)& (w_{10})\\
0.2160(26)(8)& (w_{(00)})
\end{array}\right.
\end{equation}
where the first error is
experimental (dominated by $us$ spectral errors) and the
second the nominal theoretical error. The nominal theory
error is obviously much smaller than the observed $s_0$-instability
in the $w_{(00)}$ case, and hence unrealistically small.
Comparison to the results of earlier $\Delta\Pi_\tau$ FESR 
analyses~\cite{gamizetal,gamizetalnew,kmcw,mwbrn08}
shows the significant impact of recent, improved $us$ experimental
results on the $\vert V_{us}\vert$ central values. The decreases
represented by the remeasured $us$ branching fractions, lead to
$\vert V_{us}\vert$ results $2-3\sigma$ below the 3-family-unitarity 
expectation, $0.2255(1)$~\cite{hardytowner}.

\begin{figure}
  \caption{\label{fig3} $\vert V_{us}\vert$ versus $s_0$ from
the $\Delta\Pi_\tau$ FESRs for, from top to bottom,
$w_{20}$, $\hat{w}_{10}$, $w_{10}$ and $w_{(00)}$, with the spectral
input modified by rescaling up by $3\sigma$ the branching fractions of all
modes not yet remeasured by either BaBar or Belle.}
\rotatebox{270}{\mbox{
  \begin{minipage}[t]{2.6in}
\includegraphics[width=2.55in]{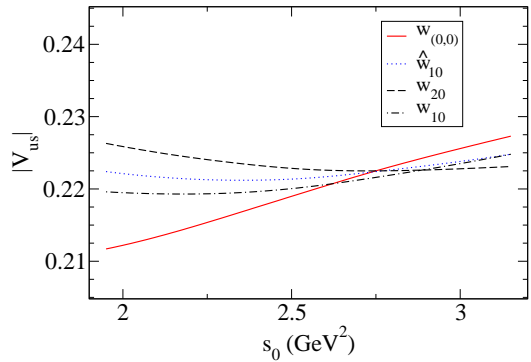}
  \end{minipage}
}}
\end{figure}

It should be stressed that several important strange decay
modes have yet to be remeasured by either BaBar or Belle, and that the
level of consistency of the $s_0=m_\tau^2$ results for different
weights could be significantly affected by such future remeasurements.
As an illustration, we show, in Figure~\ref{fig2}, the impact on
$\vert V_{us}\vert$ as a function of $s_0$ of
rescaling upward by $3\sigma$ the as-yet-unremeasured 
$\bar{K}^0\pi^-\pi^0$ branching fraction, and hence
also the $\bar{K}^0\pi^-\pi^0$ component of the $us$ spectral
distribution employed above. The issue of whether plausible shifts in 
the as-yet-unremeasured branching fractions are capable of
restoring agreement with 3-family-unitarity expectations
is less clear. In fact, it would take simultaneous $3\sigma$ upward
rescalings of all currently unremeasured $us$ branching fractions
to restore agreement.
Such a rescaling, moreover, does not produce a convincing
$s_0$-stability plateau for any of the weights considered,
as shown in Figure~\ref{fig3}. 

\begin{figure}
   \caption{\label{fig4}$\vert V_{us}\vert$ as a function of $s_0$ from
the mixed $\tau$-electroproduction FESRs for, from top to bottom
at the left, $\hat{w}_{10}$, $w_{(00)}$, $w_3$, $w_4$ and $w_2$.}
\rotatebox{270}{\mbox{
  \begin{minipage}[t]{2.6in}
\includegraphics[width=2.5in]{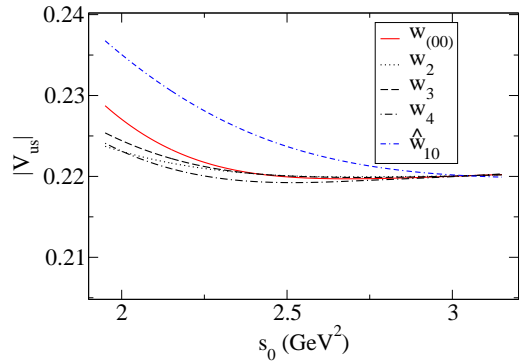}
  \end{minipage}
}}
\end{figure}

The results for $\vert V_{us}\vert$ for the $\Delta\Pi_M$ FESRs 
based on $w_{(00)}(y)$, the weight $\hat{w}_{10}(y)$ displaying the best 
$s_0$-stability for the $\Delta\Pi_\tau$ FESR, and the weights $w_2$, 
$w_3$ and $w_4$, where $w_N = 1-{\frac{N}{N-1}}y+{\frac{y^N}{N-1}}$,
are displayed in Fig.~\ref{fig4}. The weight $w_N$ produces a single
surviving integrated $D=2N+2>4$ OPE contribution 
suppressed by the coefficient $1/(N-1)$ and scaling
as $1/s_0^N$, making it a useful choice in this case,
where the slow integrated $D=2$ convergence found for the
$w_N$ versions of the $\Delta\Pi_\tau$ FESRs is no longer relevant.

If it was poor $D=2$ convergence which was responsible for the
$s_0$-instability of the $w_{(00)}$ $\Delta\Pi_\tau$ FESR results,
one would expect to see a much improved stability plateau for
the corresponding $\Delta\Pi_M$ FESR, as is indeed
found. The very good stability for the $w_N$ results
also indicates that the integrated $D=2N+2$ contributions relevant
to these cases become negligible in the upper part of
the $s_0$ window displayed in the Figure. Since, however,
$D\geq 6$ contributions increase in going from $\Delta\Pi_\tau$
to $\Delta\Pi_M$~\cite{kmtauem08}, one would expect
the instability for weights like $\hat{w}_{10}$, which do
not suppress these to the same extent as do the other weights
considered, to be enhanced, as is indeed found to
be the case. Even so, the $\hat{w}_{10}$ results converge
well to the stable results for the other weights as
$s_0\rightarrow m_\tau^2$. 

Given the very good stability
of the $w_{(00)}$ results, it is possible to quote 
a final result based on the $s_0=m_\tau^2$ version of the $w_{(00)}$ FESR,
which allows us to take advantage of the improvements in
the $us$ branching fraction errors. The result is
\begin{equation}
\vert V_{us}\vert =0.2208(27)(28)(5)(2)
\end{equation}
where the first three errors are due to the 
uncertainties on the $us$ V+A, residual $I=0$ EM  and residual $ud$ V/A 
spectral integrals, respectively, and the fourth is due to the
$D=2$ and $4$ OPE uncertainties (see Ref.~\cite{kmtauem08} for
further details). 

We conclude by stressing that, for both the $\Delta\Pi_\tau$ and
$\Delta\Pi_M$ FESRs, improved errors on $dR_{V+A;us}/ds$ are crucial.
This requires both remeasurements
of as-yet-unremeasured strange mode branching fractions, 
pursuit of higher multiplicity modes with branching fractions
down to the few$\times 10^{-5}$ level, and, in particular, a full
investigation of the $K 3\pi$ and $K 4\pi$ modes, which were
not in fact measured, but rather estimated, in the earlier experimental
analyses.

\end{document}